\documentclass[onecolumn,tighten]{aastex631}
\pdfoutput=1
\usepackage{CJK}

\usepackage{enumitem}
\usepackage{xspace}
\usepackage[multiple]{footmisc}

\usepackage{hyperref}
\usepackage{bbding}

\usepackage{comment}

\newcommand{\cntext}[1]{\begin{CJK*}{UTF8}{gbsn}#1\end{CJK*}}

\setlist{noitemsep} 

\def\m87{M87$^*$\xspace}
\def\sgra{Sgr~A$^*$\xspace}

\def\lsim{\mathrel{\raise.3ex\hbox{$<$\kern-.75em\lower1ex\hbox{$\sim$}}}}
\def\gsim{\mathrel{\raise.3ex\hbox{$>$\kern-.75em\lower1ex\hbox{$\sim$}}}}
\defcitealias{Issaoun_2022}{Issaoun, Wielgus et al. 2022}
\defcitealias{Jorstad_2023}{Jorstad, Wielgus et al. 2023}
\defcitealias{Crew_2023}{Crew et al. 2023}
\defcitealias{EHT_345}{Raymond, Doeleman et al. 2024}

\definecolor{darkred}{rgb}{0.55, 0.0, 0.0}
\definecolor{lightgray}{rgb}{0.6, 0.6, 0.6}
\definecolor{amethyst}{rgb}{0.6, 0.4, 0.8}

\begin{document}

\title{\large Mid-Range Science Objectives for the Event Horizon Telescope}

\collaboration{0}{The Event Horizon Telescope Collaboration}
\author[0000-0002-9475-4254]{Kazunori Akiyama}
\affiliation{Massachusetts Institute of Technology Haystack Observatory, 99 Millstone Road, Westford, MA 01886, USA}
\affiliation{National Astronomical Observatory of Japan, 2-21-1 Osawa, Mitaka, Tokyo 181-8588, Japan}
\affiliation{Black Hole Initiative at Harvard University, 20 Garden Street, Cambridge, MA 02138, USA}

\author[0000-0002-7816-6401]{Ezequiel Albentosa-Ruíz}
\affiliation{Departament d'Astronomia i Astrofísica, Universitat de València, C. Dr. Moliner 50, E-46100 Burjassot, València, Spain}

\author[0000-0002-9371-1033]{Antxon Alberdi}
\affiliation{Instituto de Astrofísica de Andalucía-CSIC, Glorieta de la Astronomía s/n, E-18008 Granada, Spain}

\author{Walter Alef}
\affiliation{Max-Planck-Institut für Radioastronomie, Auf dem Hügel 69, D-53121 Bonn, Germany}

\author[0000-0001-6993-1696]{Juan Carlos Algaba}
\affiliation{Department of Physics, Faculty of Science, Universiti Malaya, 50603 Kuala Lumpur, Malaysia}

\author[0000-0003-3457-7660]{Richard Anantua}
\affiliation{Black Hole Initiative at Harvard University, 20 Garden Street, Cambridge, MA 02138, USA}
\affiliation{Center for Astrophysics $|$ Harvard \& Smithsonian, 60 Garden Street, Cambridge, MA 02138, USA}
\affiliation{Department of Physics \& Astronomy, The University of Texas at San Antonio, One UTSA Circle, San Antonio, TX 78249, USA}

\author[0000-0001-6988-8763]{Keiichi Asada}
\affiliation{Institute of Astronomy and Astrophysics, Academia Sinica, 11F of Astronomy-Mathematics Building, AS/NTU No. 1, Sec. 4, Roosevelt Rd., Taipei 106216, Taiwan, R.O.C.}

\author[0000-0002-2200-5393]{Rebecca Azulay}
\affiliation{Departament d'Astronomia i Astrofísica, Universitat de València, C. Dr. Moliner 50, E-46100 Burjassot, València, Spain}
\affiliation{Observatori Astronòmic, Universitat de València, C. Catedrático José Beltrán 2, E-46980 Paterna, València, Spain}
\affiliation{Max-Planck-Institut für Radioastronomie, Auf dem Hügel 69, D-53121 Bonn, Germany}

\author[0000-0002-7722-8412]{Uwe Bach}
\affiliation{Max-Planck-Institut für Radioastronomie, Auf dem Hügel 69, D-53121 Bonn, Germany}

\author[0000-0003-3090-3975]{Anne-Kathrin Baczko}
\affiliation{Department of Space, Earth and Environment, Chalmers University of Technology, Onsala Space Observatory, SE-43992 Onsala, Sweden}
\affiliation{Max-Planck-Institut für Radioastronomie, Auf dem Hügel 69, D-53121 Bonn, Germany}

% ...

\begin{abstract}
The first images of the black holes in \sgra\ and \m87 have created a wide range of new scientific opportunities in gravitational physics, compact objects, and relativistic astrophysics.  We discuss here the scientific opportunities that arise from the rich data sets that have already been obtained and the new data sets that will be obtained, exploiting a wide range of technical advances, including observational agility, receiver upgrades, and the addition of new stations.  This document provides a 5-year framework for Event Horizon Telescope (EHT) science structured around four fundamental questions that are used to prioritize the analysis of existing data, guide technical upgrades, and determine the optimal use of future observational opportunities with EHT, ALMA, and multi-wavelength facilities. Through enhancements over this period, the EHT will create the first movie of \m87 connecting black hole and jet physics, provide detailed studies of the structure and dynamics of \sgra, characterize the magnetospheres of both systems through polarimetric imaging, and explore the spacetime properties of black holes with greater precision and range.
\end{abstract}

\date{\today}

\section{Introduction}

The EHT Collaboration (EHTC) has published total intensity and polarized images of the supermassive black holes (SMBHs) in \m87 \citep{PaperI,PaperII,PaperIII,PaperIV,PaperV,PaperVI,PaperVII,PaperVIII,2023ApJ...957L..20E,2024A&A...681A..79E} and \sgra \citep{SgrAEHTCI,SgrAEHTCII,SgrAEHTCIII,SgrAEHTCIV,SgrAEHTCV,SgrAEHTCVI,SgrAEHTCVII,SgrAEHTCVIII}, providing the first direct visual evidence for the existence of black holes as predicted by general relativity. These images have enabled  precise mass measurements for \m87, resolving a discrepancy between stellar and gas-dynamics based mass estimates, and confirmed the stellar orbit mass measurements for \sgra. Additionally, these observations have constrained the black hole inclination, plasma parameters, and magnetic field structure. 
These results have provided a powerful new laboratory to study both gravitational physics and the accretion environment for SMBHs in the context of extensive multi-wavelength campaigns on these objects \citep{EHT_MWL}. They have also enabled the highest resolution studies of several bright AGN, including the nearby radio galaxies Centaurus~A \citep{Janssen_2021} and 3C 84 \citep{2024A&A...682L...3P} and the powerful blazars 3C279 \citep{Kim_2020}, J1924-2914 \citep{Issaoun_2022}, and NRAO\,530 \citep{Jorstad_2023}.  In total, the EHTC has published over 200 refereed papers on observational, technical, and theoretical topics with over 14,000 citations in the past 5 years.\footnote{\url{https://ui.adsabs.harvard.edu/public-libraries/VFaebic8RGqFB9UuW9corg}}

\begin{figure}[tb]
    \centering
    \includegraphics[width=0.75\textwidth]{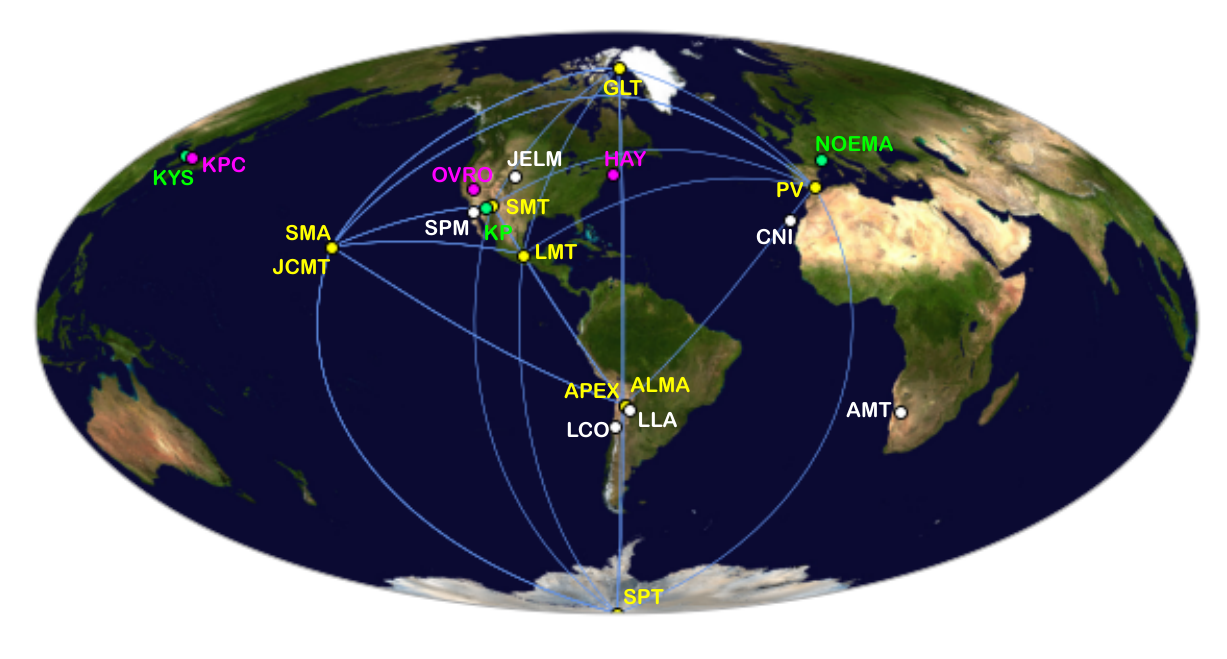}
    \caption{Map of the EHT. Stations active in 2017 and 2018 are shown with connecting lines and labeled in yellow, sites that joined in 2021-2024 are labeled in green, sites expected to join in the next 1-2 years are shown in magenta, sites expected to join over the next 5 years are shown in white. These future additions include the 37-m Haystack Telescope \citep{Kauffmann_2023}, 
    a 10.4-m telescope at the Owens Valley Radio Observatory in California,
    and the 21-m Pyeongchang
    Radio Observatory of the Korean VLBI Network \citep{Asada_2017}. 
    Other potential new sites include the 15-m Africa Millimetre Telescope \citep{Backes_2016}, the 12-m Large Latin American Millimeter Array \citep{Romero_2020}, and several new antennas proposed to be deployed through the ngEHT program
    \citep[Canary Islands, Spain; Jelm Observatory, USA; San Pedro Martir, Mexico; and Los Cumbres Observatory, Chile; ][]{2023Galax..11..107D}.
    }
    \label{fig:eht_globe}
\end{figure}

At present, all published EHT results use data from the 2017 and 2018 observing campaigns. The EHT has the opportunity to significantly improve upon its published results through the analysis of more recent data from observing campaigns in 2018, 2021, 2022, 2023, and 2024 (summarized in \autoref{tab:eht_observations}). These campaigns have already expanded to new sites (GLT, KP, and NOEMA; see \autoref{fig:eht_globe}), have doubled the recorded bandwidth (to $64\,{\rm Gb/s}$), and have demonstrated successful observations at $345\,{\rm GHz}$ \citep{Crew_2023,EHT_345}, which was utilized as a standard EHT observing frequency for the first time in the 2023 campaign. Beyond the opportunities available through these existing observations, the EHT has the potential to deliver breakthrough science over the next 5 years ($2024-2029$) by pursuing a series of technical upgrades and expansions that are coordinated with increasingly ambitious observations. 

In this paper, we describe the major mid-range scientific opportunities for the EHT and the developments that the EHTC is pursuing to achieve them. This is a living document that we expect to update as our strategies evolve with new scientific discoveries and
technical capabilities.
While the EHTC will continue to produce major results on non-horizon targets obtained as calibrators or PI-led, we focus only on \m87 and \sgra because they define the collaboration outputs. In particular, we have identified several key science goals in the mid-term, each of which is enabled by a corresponding  set of observational advances\vspace{-0.1cm}  
\begin{enumerate}
    \item {\bf Science Question:} What is the origin of relativistic jets from SMBHs?\\
    {\bf Observational Goal:} Establish the dynamical relationship between the SMBH and its relativistic jet in \m87.\\
          {\bf Requirements:} Improved baseline coverage, especially on scales of $100-500\,\mu{\rm as}$ ($0.4 - 2\,{\rm G}\lambda$), and temporal coverage extending over at least $1-3$\,months.\vspace{0.1cm}
    \item {\bf Science Question:} What causes flaring near SMBHs?\\
    {\bf Observational Goal:} Measure the ring structure and dynamics of \sgra, especially during multi-wavelength flares.\\
          {\bf Requirements:} Improved snapshot baseline coverage, with mutual visibility from $6-8$ separate sites and coordinated multi-wavelength observations in the infrared and X-rays.\vspace{0.1cm}
    \item {\bf Science Question:} Do SMBHs have strong magnetospheres that extract their spin energy?\\
    {\bf Observational Goal:} Measure the near-horizon magnetic field structure and strength in \m87 and \sgra.\\
          {\bf Requirements:} Full-Stokes imaging of linear polarization, rotation measure, and circular polarization combining separate sidebands and $230+345\,{\rm GHz}$ observations.\vspace{0.1cm}    
    \item {\bf Science Question:} What are the characteristics of the massive compact objects in galactic nuclei? \\
    {\bf Observational Goal:} 
    Measure properties of the emission rings, apparent shadows, and other observables that depend on the spacetime properties for both \m87 or \sgra.\\
          {\bf Requirements:} Full-Stokes, time-averaged imaging at $230$ and $345\,{\rm GHz}$, with at least 10 epochs separated by $10 GM/c^3$ (minutes for \sgra; days for \m87) and extending over a total time span of at least $1000 GM/c^3$ (6 hours for \sgra; 1 year for \m87).\vspace{-0.1cm}  
\end{enumerate}

These goals are enabled by improvements along several axes: image fidelity, time domain, multi-frequency VLBI observations, and broadband multi-wavelength observations. In \autoref{sec:development}, we describe planned improvements to the EHT. 
In \autoref{sec:eht_science}, we describe the driving scientific goals in more detail. In \autoref{sec:recommendations}, we summarize the opportunities for major scientific breakthroughs with the EHT in the coming years.

\section{Planned EHT Developments}
\label{sec:development}

Technical developments for the EHT are underway and/or planned in multiple areas.  The science impact of these developments will be both during and after the 5-year timescale of this science plan, dependent upon the time necessary to implement, commission, and exploit new capabilities.
\begin{itemize}
    \item {\bf Pipeline software:} The EHT makes use of 
    dedicated software pipelines and libraries for processing its unique data from the observatory 
    \citep{blackburn_2019, janssen_2019}. There are several efforts underway to standardize and modernize EHT processing and analysis tools to further automate workflows and ultimately reduce the time required to produce science \citep[e.g.,][]{2022Galax..10..119H, Tiede_2022,vanBemmel_2022}.
    \item {\bf Agility:}  Technological and operational advances are underway to simplify observations and enable more remote control.  Improved observing agility will enable movie campaigns, better response to weather, and response to transient events.
    \item {\bf Receiver upgrades:}  The EHT is pursuing development of dual- and tri-band receivers covering ALMA bands 3, 6, and 7.  While observations have been successfully conducted in each band separately, sensitivity and atmospheric coherence time decreases at shorter wavelengths.  This can be addressed through simultaneous observations in two or more bands and application of the frequency-phase transfer technique to extend coherence times at short wavelengths and, therefore, improve sensitivity \citep{2023arXiv230604516D,galaxies11010016,Pesce_2024}.
    Bandwidths in each receiver will also be doubled to match the ALMA Wideband Sensitivity Upgrade \citep{2023pcsf.conf..304C}.  Shorter wavelength VLBI, i.e. in ALMA Band 9, has been proposed \citep[e.g.,][]{2023PASP..135i5001C} but is not yet a part of the EHT plan.
    \item {\bf Bandwidth upgrade:} New developments in digitization, digital signal processing, and high bandwidth recording are necessary to accommodate the increase in number of receivers and broader bandwidths \citep{2020PASP..132h5001J}.
    \item {\bf Upgraded and new stations:} Developments are underway with antennas at Haystack Observatory, Owens Valley Radio Observatory, and the Korean VLBI Network to enable their participation in upcoming EHT campaigns \citep{2023Galax..11....9K,2022JKAS...55..207S}.  Design and development of the Africa Millimeter Telescope in Namibia is underway \citep{Backes_2016,2023A&A...672A..16L}.  Relocation of the Greenland Telescope (GLT) to the summit of Greenland \citep{2023PASP..135i5001C} and construction of an array of new 9--m dishes as part of the ngEHT project \citep{2023Galax..11..107D} have been proposed. One longer term vision for an expanded array, including some of the above facilities, has been developed through the ngEHT program and published in a special issue of the  journal Galaxies (\url{https://www.mdpi.com/journal/galaxies/special_issues/ngEHT_blackholes}). New stations will improve uv-coverage, array sensitivity, and robustness to station-loss due to weather.
    \item{{\bf Spectral line observations:} Spectral line observations provide information on the kinematics, physical and chemical composition of the gas in astronomical objects. Spectral line VLBI with the EHT+ALMA can support core black-hole studies via, e.g., high resolution observations of water megamaser at 321\,GHz and 325\,GHz\footnote{It is worth noting that more water maser lines are known to emit from AGN disks, e.g. at 183\,GHz, but suitable receivers are not uniformly available on all EHT stations.} \citep{Gray2016}. Absorption studies offer another powerful tool to study and resolve the molecular gas in the vicinity of AGN, allowing to detect diffuse gas even at high redshift, depending only on the background continuum emission and the gas optical depth   
    \citep[see e.g.,][]{2023Galax..11...10K}.

   }
    
\end{itemize}

Similar technical developments are also discussed in \cite{2023Galax..11..107D} which presents a reference array and design considerations from the ngEHT project.

\begin{figure}[t]
    \centering
    \includegraphics[width=0.45\textwidth]{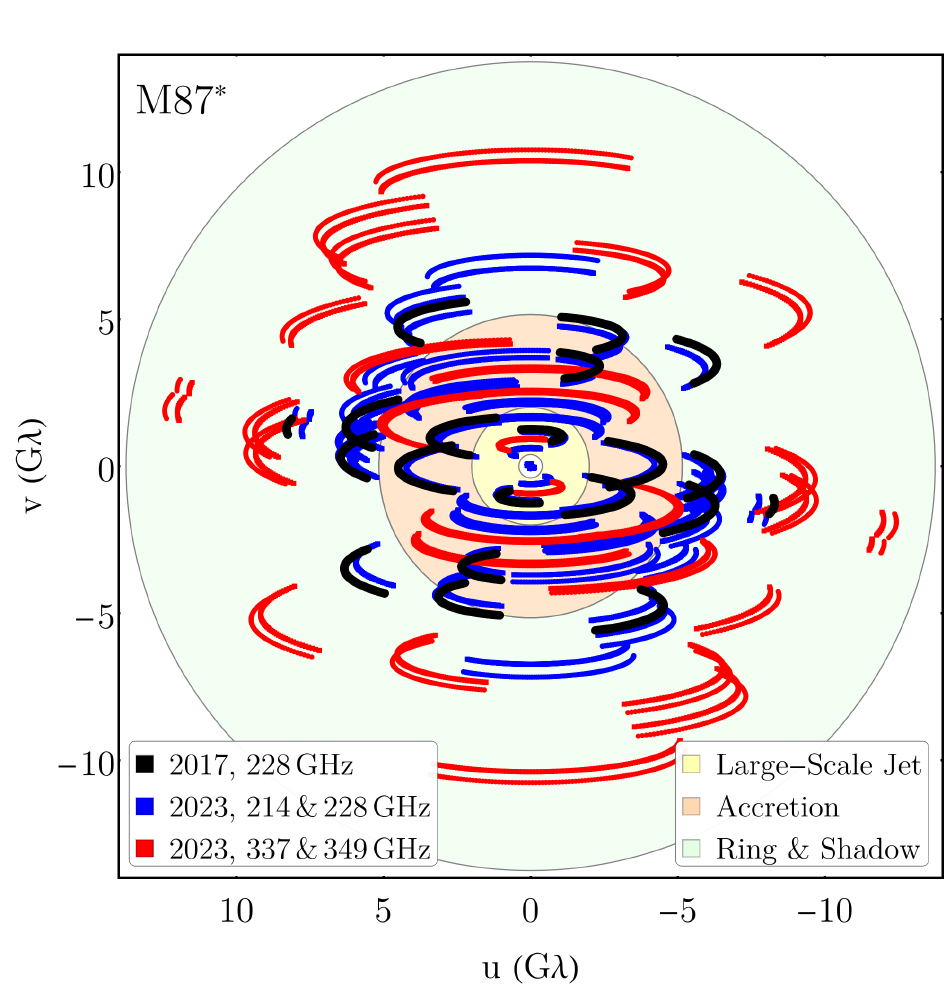}
    \includegraphics[width=0.45\textwidth]{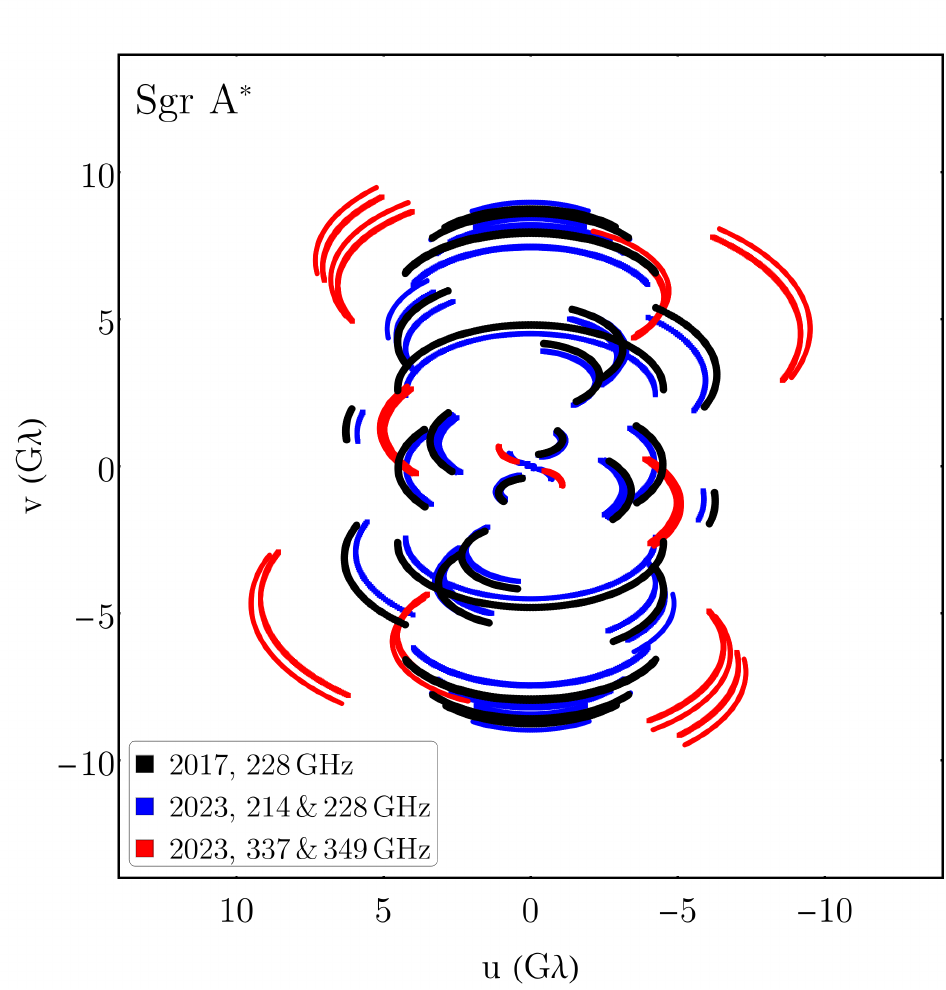}
    \caption{Comparison of EHT geometric baseline coverage over time for \m87 (left) and \sgra (right). Black curves show the maximal coverage with the 2017 EHT array; blue/red curves show the dual-sideband $214/228\,{\rm GHz}$ and $337/349\,{\rm GHz}$ coverage for the 2023 EHT array. All current EHT sites have a telescope with $345\,{\rm GHz}$ observing capabilities except SPT, Kitt Peak, and LMT. For the \m87 coverage, baseline ranges that contribute to the various science goals are colored: scales of $10-40\,\mu{\rm as}$ probe the emission ring and apparent shadow, $40-100\,\mu{\rm as}$ probe the accretion and jet launching regimes, and $100-500\,\mu{\rm as}$ probe the large-scale jet. These figures show the maximal baseline coverage of each array and do not account for detection limitations from finite sensitivity, which can  limit the accessible baseline coverage at 345~GHz \citep[see \autoref{fig:M87_imaging}; ][]{Pesce_2024}.
    }
    \label{fig:uv}
\end{figure}


\section{Science Opportunities}
\label{sec:eht_science}

\subsection{Establish the dynamical relationship between the SMBH and its relativistic jet in \m87}

Black holes impart variability on their surrounding environment through channels including circularization of accreting material and frame dragging near a spinning black hole. One of the most exciting (but untested) predictions from current EHT simulations is that the relativistic jet in \m87 is powered by drawing its energy from the spin of the SMBH through an analogue of the Penrose process \citep{Penrose_1969,Blandford_Znajek_1977,PaperV,PaperVIII}. 
This process is considered an extension of the concept initially proposed by Penrose, where energy is extracted from a rotating black hole. In the Blandford-Znajek process, this energy extraction is facilitated by magnetic fields interacting with the black hole's spin, leading to the formation of relativistic jets.
In addition, major unresolved questions of particle heating and jet formation have distinctive corresponding dynamical signatures. The jet structure and dynamics of \m87 have been extensively studied on scales of 10-1000 Schwarzschild radii using VLBI at frequencies up to 86\,GHz  \citep[e.g.,][]{Kovalev_2007,Mertens_2016,Hada_2016,Walker_2018,Kim_2018,2023Natur.616..686L,2023Natur.621..711C}. These studies have revealed a relativistic outflow with substantial evolution on short (${\sim}$days) timescales and a wobbling position angle on long (${\sim}$years) timescales. However, these dynamics have never been directly connected to the black hole, and the EHT has the angular resolution necessary to monitor launching and acceleration of the jet and to reveal jet-disk kinematics. 
Analysis of the near-horizon jet morphology, radial velocity, azimuthal velocity, and polarization structure over time could test the jet-launching mechanism, and constrain black hole parameters such as magnetization and black hole spin \citep{2021ApJ...914...55W,2022NatAs...6..103C,2022A&A...660A.107F,Chael_2023,2023ApJ...959L...3D,2024ApJ...960..106M}.
Hence, {\bf full-polarization dynamical studies of \m87 are a major priority for continued EHT observations.}

The relativistic jet in \m87 is associated with dynamical activity on the scales of ${\sim}$weeks, and GRMHD simulations predict that the near-horizon emission should also be variable, with an associated correlation timescale of $\sim 50-100MG/c^{3}$, or a few weeks. Currently published EHT data can only marginally constrain the structure of the extended jet emission \citep[e.g.,][]{PaperIV,Arras_2022,Broderick_2022}. In addition, while the published set of EHT observations unambiguously identifies the existence of structural evolution near the black hole \citep{2024A&A...681A..79E}, {the published EHT temporal coverage is inadequate to unambiguously identify the structural variability in \m87} (e.g., to determine apparent velocity of rotation, inflow, or outflow). 

\begin{figure}[t]
    \centering
    \includegraphics[width=\textwidth]{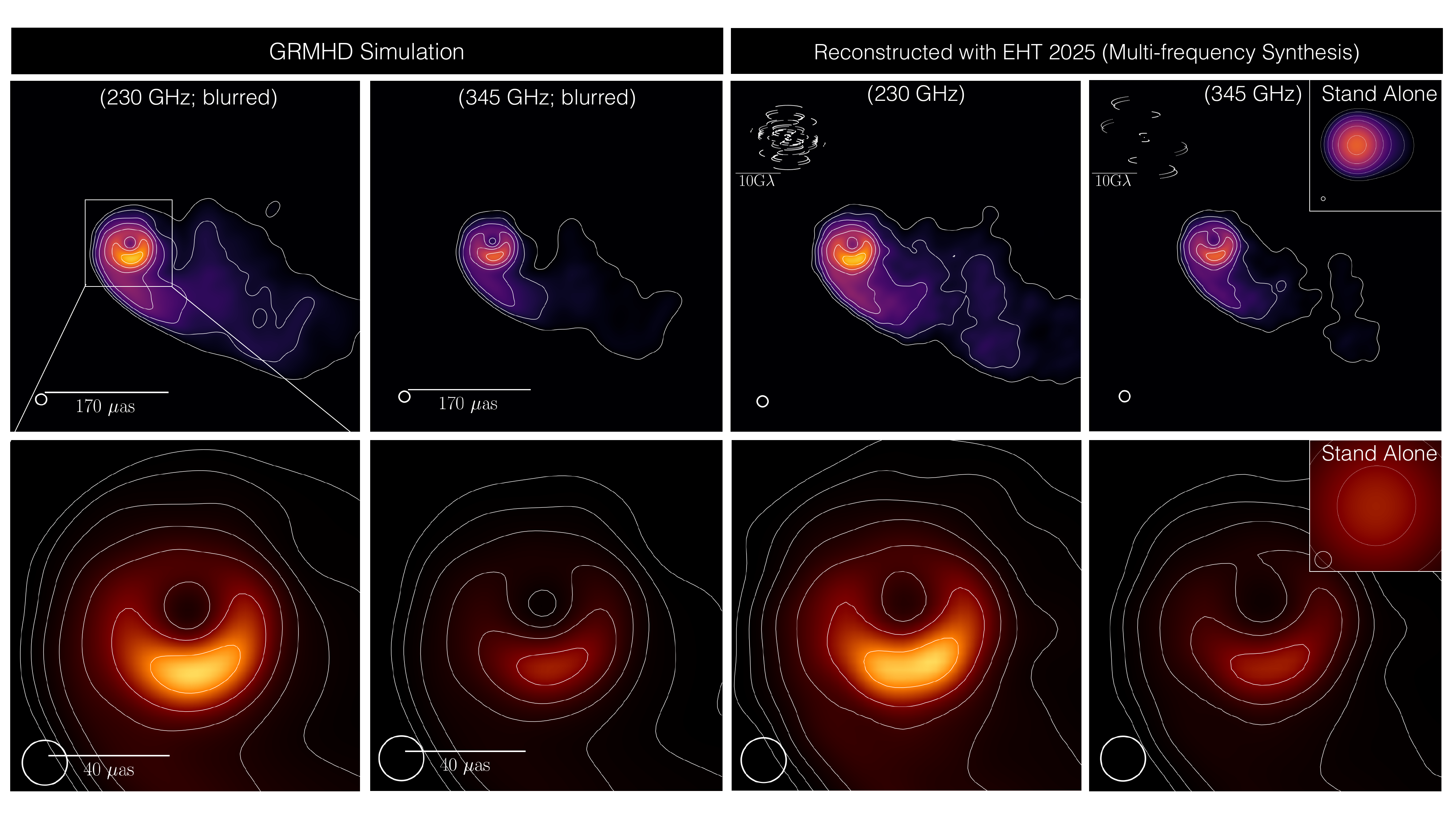}
    \caption{
(From left to right) First two columns show images of \m87 at 230 and 345\,GHz from a GRMHD simulation \citep{Chael_2019}. The next two columns show multi-frequency reconstructions \citep{Chael_2023_multifreq} using simulated data from the EHT with the projected 2025 array configuration and realistic weather modeling \citep{Pesce_2024}. Insets show the baseline coverage in each band as well as reconstructions using 345\,GHz data alone. Dual-band observations are necessary to obtain images at 345\,GHz. The top row shows large-scale emission with a logarithmic color scale; the bottom row shows the central region with a linear color scale. Contours are set by successive powers of 1/2 from the peak brightness.  Enhancements to imaging through inclusion of 345\,GHz are a strong function of the underlying source model.
    } 
    \label{fig:M87_imaging}
\end{figure}

Meaningful studies of \m87 dynamics with the EHT will require improved image dynamic range to unambiguously constrain the extended jet structure \citep[see, e.g.,][]{Johnson_2023,Conroy_2023}. This requires multiple sensitive baselines that sample angular scales of $100-500\,\mu{\rm as}$ ($|\vec{u}| = 0.4 - 2\,{\rm G}\lambda$), ideally with associated phase information (through closure phase) and full-Stokes information. For comparison, the EHT observations in 2017 included only a single intersite baseline shorter than $2\,{\rm G}\lambda$, the LMT-SMT baseline, which sampled the range $1.25 - 1.5\,{\rm G}\lambda$ (see \autoref{fig:uv}). However, the improved baseline coverage in the current EHT may allow firm detections of the \m87 jet (see \autoref{fig:M87_imaging}). To study the dynamics of \m87 will also require longer EHT observing campaigns, extending for at least $1-3$\,months with a sub-week observing cadence, and the study of observations over several years to assess the connection between the evolving jet and the horizon-scale structure. These campaigns can also be coordinated with multi-wavelength facilities to elucidate the mechanisms behind particle acceleration within the jet and the generation of high energy flares. 

\subsection{Measure the ring structure and dynamics of \sgra, especially during multi-wavelength flares}

\sgra is a highly variable source, with a gravitational timescale of only $GM/c^3 \approx 20\,{\rm seconds}$ and daily flares in the radio through X-rays \citep[e.g.,][]{Baganoffetal2001,Genzel_2003,Eckart_2004,Marrone_2008,Haggardetal2019,Witzel_2021}. The EHT is the only facility that can directly image the evolving emission structure of \sgra. 
Measuring the direction of rotation would enable the EHT to corroborate the measurement of clockwise motion by the GRAVITY collaboration \citep{2018A&A...618L..10G,2020A&A...643A..56G}. Measuring the magnitude of the apparent rotation rate and the pitch angle of orbiting features would enable constraints of inclination, distance-independent constraints on mass, and constraints of spin \citep{2022ApJ...941L..12R,Conroy_2023}.  Movies depicting the near-horizon dynamics of Sgr A* could also constrain different models of flaring events \citep{2020ApJ...900..100R,2022ApJ...924L..32R,2021MNRAS.502.2023P,2024MNRAS.530.1563G}.

However, the rapid variability renders the standard VLBI approach of Earth-rotation synthesis inapplicable, and the ``snapshot'' baseline coverage of the published EHT data are inadequate to unambiguously describe the source structure \citep{SgrAEHTCIII}. 

In the 2017 EHT campaign, the variability of \sgra in total intensity was mild on April 6 and 7 and more significant on April 11 \citep{SgrAEHTCIV,Wielgus_2022}. On all days, the polarimetric variability is considerably stronger than that in total intensity alone \citep{SgrAEHTCVII}. This is consistent with previous measurements of the 1.3 mm light curve \citep{Marrone_2006,Bower_2018,Goddi_2021} and in interferometry with a precursor EHT array \citep{Johnson_2015}.

{\bf Analysis of the structural evolution associated with the variability of \sgra is a major opportunity for the EHT in near-term analyses.} Despite the current limitations in snapshot baseline coverage, variability analysis will be carried out through non-imaging studies that quantify the power spectrum of the variability \citep[e.g.,][]{Broderick_2022} or that rely on constraining low-dimensional parametric models \citep[e.g.,][]{SgrAEHTCIII,SgrAEHTCIV,Roelofs_2023,SgrAEHTCVII}, alongside attempts at dynamic imaging with the available data.

\subsection{Measure the near-horizon magnetic field structure and strength in \m87 and \sgra}

The EHT has published findings on both the linear and circular polarization results for \m87 \citep{PaperVII,PaperVIII,2023ApJ...957L..20E} and \sgra \citep{SgrAEHTCVII,SgrAEHTCVIII}. For \m87, the addition of polarization indicated a preference for magnetically arrested accretion flow models and severely restricted the acceptable GRMHD parameter space; for instance, passing models in total intensity had accretion rates $\dot{M}$ that extended over $10^{-4} - 10^{-1} M_\odot/{\rm yr}$, while the linear polarization constraints reduced the range of passing models to $10^{-4} - 10^{-3} M_\odot/{\rm yr}$ \citep{PaperVIII}. Consistent with our analyses of \m87, the \sgra polarimetric data further supports the preference for GRMHD models characterized by dynamically significant magnetic fields \citep{SgrAEHTCIII}.
The high degree of spatially resolved linear polarization observed in \sgra \citep[ranging from 24\% to 28\%, with peaks at approximately 40\%;][]{SgrAEHTCVII} provides a stringent constraint on the parameter space, effectively ruling out models that exhibit excessive Faraday depolarization. Polarimetric variability also has the potential to characterize turbulence in the accretion flow. Thus, polarimetric information is imperative in guiding astrophysical models for \m87 and \sgra, and our current analyses have only begun to explore the information encoded in the polarized signals from these SMBHs.

In addition, while linear polarization images primarily constrain the magnetic field \emph{structure}, frequency-dependent Faraday rotation primarily constrains the magnetic field \emph{strength}. Because all the current EHT results are effectively monochromatic, they have not yet integrated information that is accessible through resolved Faraday rotation images that may be possible through dual-sideband (214/228\,GHz) observations in 2018 onward, and through dual frequency (230/345\,GHz) observations in 2023 onward. Both unresolved measurements with ALMA \citep[e.g.,][]{Bower_2018,Goddi_2021,2022A&A...665L...6W} and GRMHD simulations \citep[e.g.,][]{Moscibrodzka_2017,Ricarte_2020} indicate that the EHT observations will show rich frequency structure from internal Faraday rotation, providing crucial astrophysical insights. Moreover, measuring the Faraday rotation will be imperative so that the observed linear polarization can be de-rotated to evaluate the internal magnetic field structure. This is particularly relevant for \sgra, where the inferred sense of rotation of the material around the ring depends on whether the Faraday rotation occurs internally within the emitting plasma or in an external screen \citep{SgrAEHTCIII}.

\subsection{Estimate the spacetime properties for \m87 and \sgra through properties of their emission rings and apparent shadows}

The EHT has enabled crucial new constraints on the compact objects in \m87 and \sgra. For both sources, studies of the emission ring and apparent ``shadow'' \citep{Falcke_2000} yield estimates of the black hole mass-to-distance ratio to an accuracy of $10-20\%$ \citep{PaperVI,SgrAEHTCIV,SgrAEHTCVI}. For \m87, the EHT measurements of the SMBH mass conclusively resolved a long-standing dispute between estimates of the mass from gas dynamical and stellar dynamical measurements on scales ${\sim}10^4-10^5$ times larger than those probed by the EHT. For \sgra, the EHT measurements showed precise agreement with the measurements from resolved stellar orbits -- a powerful confirmation of general relativity. These measurements allow constraints on the spacetime parameters for various non-Kerr spacetimes as well as excluding some families of alternative compact objects \citep[e.g.,][]{Mizuno_2018,Psaltis_2020,Kocherlakota_2021,SgrAEHTCVI,Younsi_2023}.

Nevertheless, the EHT measurements of both sources have only begun to explore what is possible through VLBI. For \m87, the central brightness depression is only constrained to a contrast of $10{:}1$, our measurements have only given an upper limit on the thickness of the emission ring, and our measurements give only weak constraints on ring ellipticity \citep{Tiede_2022}. For \sgra, the central brightness depression is only constrained to a contrast of $3{:}1$, and even the most basic azimuthal structure of the emission ring (such as the position angle of peak brightness) and its geometric shape is uncertain. Most importantly, the current EHT measurements have only weakly constrained the spin of the black hole in both sources. 

\begin{figure}[t]
\centering
\includegraphics[width=\textwidth]{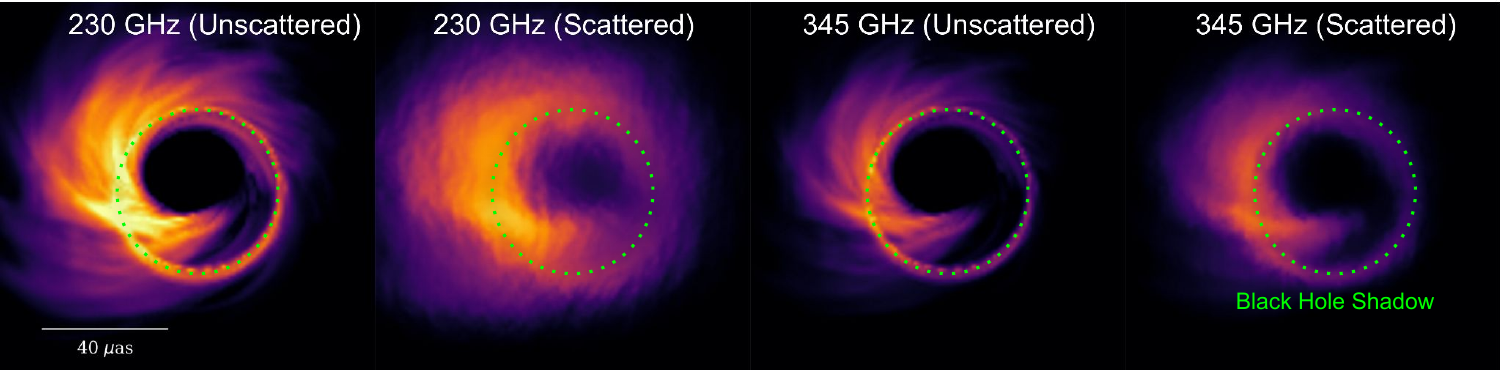}
\caption{
The effects of interstellar scattering on images of \sgra. The panels show simulated images of \sgra at 230 and 345\,GHz both before and after scattering. Even for an array with perfect imaging capabilities, 345\,GHz observations of \sgra are necessary to reduce the scattering and source opacity sufficiently to access sharp gravitational features.
}
\label{fig:scattering}
\end{figure}

The EHT has the opportunity to substantially enrich its results through improved angular resolution observations of both sources. The most straightforward pathway to sharper images is through higher frequency observations, with 345\,GHz coming online as a standard observing mode since 2023. The observations at 345\,GHz (0.8\,mm) will increase the effective resolution of the array by up to 50\%. Higher frequency observations are especially critical for \sgra because of its severe interstellar scattering, which diminishes rapidly at higher frequencies \citep[see \autoref{fig:scattering};][]{Shen_2005,Bower_2006,Psaltis_2018,Johnson_2018,Issaoun_2019}. The 345\,GHz observations will also be imperative for stronger tests of the spacetime properties through comparison with the 230\,GHz observations because the strong gravitational lensing that gives rise to features such as the black hole shadow \citep{Falcke_2000}, photon ring \citep{Johnson_2020}, and inner shadow \citep{Chael_2021} is achromatic, whereas the confounding emission from the nearby plasma is steeply chromatic. In addition to higher frequency observations, combining multiple epochs of observations of \m87 will give access to its time-averaged structure. Likewise, combining multiple epochs of observations of \sgra will provide stronger constraints on both its variability power spectrum and the time-averaged structure; these must be evaluated together because of the strong effects of intrinsic variability on the modeling and imaging \citep[e.g.,][]{SgrAEHTCIII,SgrAEHTCIV,Broderick_2022}. This structure may give crucial insights into the black hole spin through polarimetric features \citep[e.g.,][]{PWP,Moscibrodzka_2021,Ricarte_2021,Chael_2023}. In particular, the angle of linear polarization across position angle and radius has  potential to constrain spin \citep{Chael_2023}.

Valuable constraints on the spacetime properties of \sgra can also be
measured from nearby pulsars \citep[e.g.,][]{Psaltis_2016}. Expectations are that pulsars should be numerous in this region, although difficult to detect as the result of interstellar scintillation.  The EHTC has now conducted the most extensive and sensitive search for Galactic Center pulsars at millimeter wavelengths \citep{2023ApJ...959...14T}. Because pulsar searches can simply ``piggyback''
on standard EHT observations, we will continue them into the
future (also at $345\,{\rm GHz}$), as well as combining these searches with searches for Galactic Center pulsars at other frequencies.
Combining EHT's high-resolution electromagnetic observations with the rich dataset from gravitational wave observatories, such as LIGO/Virgo, LISA, or ET, enables more accurate determinations of black hole properties, such as mass, spin, and the dynamics of their surrounding accretion disks and jets. These integrated observations are pivotal for enhancing existing models and achieving unprecedented precision in our measurements. Such multi-messenger approaches are not only for corroborating the current predictions of GR but also will provide profound insights into the fundamental forces and interactions that govern our universe.

\section{Plans for Existing Data and Future Observing Campaigns}
\label{sec:recommendations}
In this section, we show how our science questions map onto analysis of existing data sets and will guide future observing campaigns.

\subsection{Opportunities with Existing EHT Data}

The EHT operations and science utilization span at a minimum a 5-year cycle, from the identification of the year-specific science goals to the completion of science utilization and the announcement of results.  Archival data is available on the EHT public website\footnote{\url{https://eventhorizontelescope.org/for-astronomers/data}}. We now briefly summarize the expected merits of unpublished EHT observations of \m87 and \sgra. \autoref{fig:eht_goals} summarizes the potential for each of these observing campaigns to address the primary goals given above. 

{\noindent \bf 2017:} Extensive analysis has been carried out on nearly all aspects of \m87 and \sgra from this campaign. One opportunity that is still under investigation is studies of \sgra dynamics during X-ray flaring activity on 11 April 2017 \citep{SgrAEHTCII}, which also shows associated polarimetric variability in the EHT observations \citep{2022A&A...665L...6W,SgrAEHTCVII}. 

{\noindent \bf 2018:} EHT observations in 2018 provide an incremental technical improvement compared to 2017. The EHTC was enhanced through the inclusion of the GLT for the first time and doubling of the recording bandwidth across all stations.  For \m87, we have confirmed the stability of the ring diameter over a one year period while also observing a change in the orientation of the azimuthal brightness distribution \citep{2024A&A...681A..79E}. For \sgra, the baseline coverage is unchanged in 2018 (the GLT cannot observe \sgra), but these observations will determine the stability of the image structure over 1-year timescales. In addition, the recorded bandwidth doubled in 2018 through dual-sideband data (from $4\,{\rm GHz}$ to $8\,{\rm GHz}$), resulting in an increase in the spanned bandwidth from $4\,{\rm GHz}$  to $18\,{\rm GHz}$. Because Faraday rotation is proportional to the squared observing wavelength, the effects of rotation measure in 2018 data will be approximately $5{\times}$ those in 2017 data. 

{\noindent \bf 2021:} EHT observations in 2021 have significantly improved baseline coverage relative to 2017. Despite the notable omission of the LMT, the number of non-redundant baselines \emph{doubles} in 2021 relative to 2017, increasing from 10 to 21 baselines for \m87 and from 15 to 21 baselines for \sgra. The 2021 observations also include new short baselines (KP-SMT and PV-NOEMA), which will improve the EHT's sensitivity to extended jet or disk structures in \m87 and \sgra. This EHT campaign also included 345\,GHz observations of \m87 on an adjacent day to 230\,GHz observations, creating the possibility to combine them for multi-frequency synthesis with up to 50\% sharper resolution than the current EHT images. For \sgra, the best \emph{snapshot} baseline coverage does not improve in 2021 because the addition of Kitt~Peak is offset by the loss of the LMT in the region of time with the best baseline coverage (and NOEMA does not have mutual visibility in this interval). 

{\noindent \bf 2022:} EHT observations in 2022 created significant new opportunities. For both \m87 and \sgra, there are 8 participating sites, giving 28 baselines. Moreover, the snapshot baseline coverage for \sgra increases from at most 5 sites (10 baselines) in 2017-2021 to 6 sites (15 baselines) in 2022, improving the capacity to constrain the time-variable structure. A limitation of the 2022 observations is that they sample only 3 days in a 9-day interval for \m87 (observations in 2021 sampled 6 days in a 10-day interval). 

{\noindent \bf 2023:} EHT observations in 2023 included for the first time data obtained at 345\,GHz on \sgra with 5 stations at three geographic locations, Chile, Hawaii, and Arizona.  These data will provide the first measurements of the angular size and asymmetry of \sgra at these wavelengths in total and polarized intensity.  Direct comparison with the 230\,GHz epochs from the same campaign will enable construction of a multi-frequency synthesis image.  These data also provide the opportunity for the first ALMA 345\,GHz search for Galactic Center pulsars.  \m87 was not observed in this epoch.

{\noindent \bf 2024:} EHT observations in 2024 included both \sgra and \m87 in three different frequency bands centered at 230, 260, and 345 GHz, an array with 11 sites including a limited first-time participation of the Korean VLBI Network (KVN) Yonsei antenna (KYS) on a best-efforts basis, and a significant multi-wavelength campaign.

\begin{table}[t]
    \centering
    \includegraphics[width=1.0\textwidth]{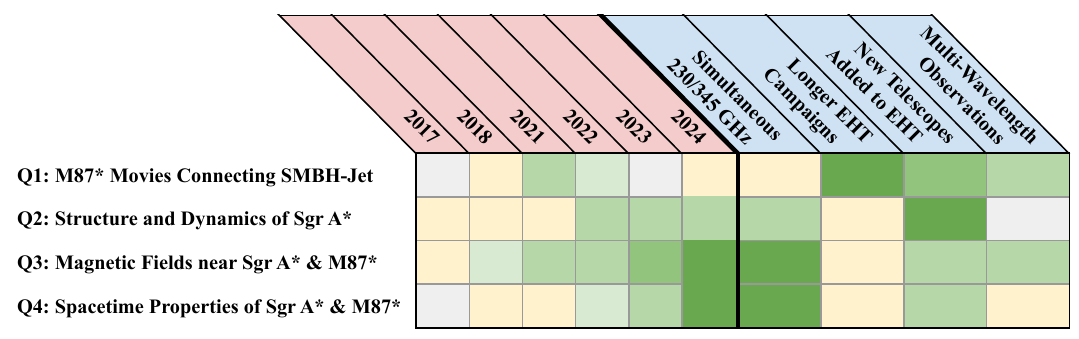}
    \caption{Summary of the primary EHT goals (rows) and the pathway to achieving them (columns). Red headers denote past or scheduled EHT campaigns (see \autoref{tab:eht_observations}); blue headers denote potential EHT upgrades (see \autoref{sec:upgrades}). Cells are colored by the likelihood to significantly advance progress on the corresponding science goal (relative to currently published results) through the associated observation or upgrade: gray indicates no advance, yellow is a minor advance, and green is a major advance shaded by significance. Dark green indicates the expectation for a decisive discovery enabled by the corresponding upgrade. 
    }
    \label{fig:eht_goals}
\end{table}

\subsection{Priorities for Mid-Range Science Campaigns and Upgrades}
\label{sec:upgrades}


In the mid-range ($2024-2029$), the EHT has many potential observations, upgrades, and expansions that can be considered. The most significant improvements will likely come in the following areas:
\begin{itemize}

\item {\bf Improved image fidelity and sensitivity through simultaneous multi-frequency observations and the use of frequency phase transfer techniques.}
The most significant recent new EHT capability is the addition of 345\,GHz as a standard observing frequency in 2023 although observations at this frequency are likely to remain limited in both sensitivity and baseline coverage. 345\,GHz will be most effective when combined with VLBI observations at other wavelengths through simultaneous dual-frequency 230+345\,GHz or triple-frequency 86+230+345\,GHz observations. Such observations will enable multi-frequency synthesis as discussed below and, significantly, will also allow the use of frequency phase transfer (FPT) techniques that exploit the fact that atmospheric phase fluctuations at these frequencies are dominated by the troposphere, which is non-dispersive. Hence, phase fluctuations can be tracked at lower frequencies and then corrected at higher frequencies where e.g. phase wrapping can become problematic. This technique has already been demonstrated at frequencies up to 130\,GHz using the KVN, extending coherence time from tens of \emph{seconds} to tens of \emph{minutes} \citep{Rioja_2015}. Lower-altitude non-345\,GHz EHT telescopes will benefit from dual-frequency FPT from 86 to 230\,GHz allowing also continued operation under less-than-ideal conditions. High-altitude/low-opacity EHT telescope will benefit from dual-frequency 230 to 345\,GHz or triple-frequency FPT to improve the coherence of VLBI at the higher band(s).
It is likely that the full angular resolution improvement possible at 345\,GHz will require this technique to offset the severe limitations from reduced sensitivity and rapid atmospheric phase fluctuations \citep{Issaoun_2023,Pesce_2024}. These improvements will be necessary to mitigate the strong interstellar scattering of \sgra and to use this source to obtain tighter constraints on potential deviations from the Kerr metric (leveraging the precise measurements of mass and distance from resolved stellar orbits), and they will enable spectral studies of the emission near \m87 that can clarify what image features are associated with strong gravitational lensing (which is achromatic). The potential merits of FPT for 230 and 345\,GHz VLBI have been extensively explored and simulated (through the ngEHT program), and the first FPT experiment between 86 and 230\,GHz took place in November~2022 (led by the KVN).\footnote{In 2022, a dedicated workshop exploring multi-band capabilities for VLBI was convened, with most presentations available online (\url{https://www.ngeht.org/broadening-horizons-2022}).}

\item {\bf Longer EHT observing windows enabled through agile observations, to produce horizon-scale movies of \m87.} Even through 2024, EHT campaigns of \m87 are still effectively within a ``snapshot'' regime. The source coherence timescale is expected to be $50-100~GM/c^3 \approx 20-40~{\rm days}$, while the longest span of existing EHT observations is only 10~days. An EHT campaign extending over 2-3 months, even with a limited observing cadence, would have immense value in constraining the source dynamics. 
It may be possible to study ring dynamics with relatively sparse uv-coverage but more detailed studies of disk-jet dynamics may require long-duration, high-cadence studies.  Detailed trade-off studies are required to optimize movie observing campaigns with scientific goals.
\item {\bf Improved dynamic range and snapshot coverage through the integration of new telescopes.} Many scientific objectives with submillimeter VLBI are simply not possible without improved baseline coverage. Additional baseline coverage is necessary to produce movies of \sgra, and to detect the jet and counter-jet in \m87. 
Some existing sites could potentially join EHT observations \citep[e.g., the 21-m Pyeongchang Radio Observatory;][]{Asada_2017} and several are being developed through the ngEHT program \citep[e.g., the Haystack 37-m and the OVRO 10.4-m;][]{Astro2020,Raymond_2021,2023Galax..11..107D}. Other potential forthcoming sites include the 15-m Africa Millimetre Telescope \citep{Backes_2016} and the 12-m Large Latin American Millimeter Array \citep{Romero_2020}.
\item {\bf Coordinated multi-wavelength observations to enable multi-frequency synthesis.} For \m87, quasi-simultaneous observations with a global 86\,GHz array (the GMVA or VLBA) would enable multi-frequency synthesis across an order of magnitude wider bandwidth than is currently sampled by the EHT. 
Moreover, it would allow the EHT to leverage the substantially higher image dynamic range at 86\,GHz, 1-2 orders of magnitude better than the EHT alone \citep[e.g.,][]{Kim_2018,2023Natur.616..686L,2023Natur.621..711C}, to improve the fidelity of EHT images \citep[see, e.g.,][]{Chael_2023_multifreq}. For \m87, these observations would ideally be scheduled within 1-2 days of each other to ensure that the source was effectively static. For \sgra, coordinated simultaneous 86\,GHz observations can provide constraints on the severe interstellar scattering as well as characterize multi-frequency light-curves during flares, which are an important aspect of EHT multi-frequency analyses. 
\item {\bf Expansion of the sample of horizon-scale objects.}  Observations are already underway to detect and characterize objects with predicted photon ring diameters that are a factor of a few smaller than that of \sgra and \m87 \citep{2023Galax..11...15R}.  Given the large systematic uncertainty in black hole masses, the photon rings in these objects are potentially resolvable with the current EHT at 230 or 345 GHz.  Characterization of these objects also explores accretion flows and jet formation on compact scales as well as establishes their suitability for photon ring imaging through high dynamic range imaging,  higher frequency VLBI, and space VLBI
\citep[e.g.,][]{2021ApJ...923..260P,2022Galax..10..109P,2023Galax..11....6R}.
\end{itemize}

We will prioritize efforts and funding that are in support of these objectives. 
Additional areas of development such as observations at frequencies higher than 345\,GHz and real-time data transport and correlation offer further possibilities but should not be pursued at the expense of these primary goals.

While this document has focused on science objectives related to \m87 and \sgra, these array enhancements would also significantly improve our understanding of relativistic jets from AGN, particularly in terms of jet formation, collimation, and acceleration, the role of magnetic fields in jet dynamics, and the processes and locations of high energy emission production. Currently, EHT observing campaigns are limited to capturing snapshot images of blazar jets with the highest possible angular resolutions. The EHT's imaging of the blazar 3C279 has uncovered an unexpected, twisted, and bent jet structure near the central black hole \citep{Kim_2020}. Similarly, observations of Centaurus A have provided detailed insights into jet collimation and its relationship with the supermassive black hole \citep{Janssen_2021}, advancing our understanding of how jets form and accelerate. These findings underscore the EHT’s transformative impact on our comprehension of relativistic jets in AGN, and while these initial results from the EHT are groundbreaking, the potential for further advancements is substantial. The planned longer observing windows will enable the creation of the first movies of blazar jets at the remarkable 20 microarcsecond resolution of the EHT. This enhancement will provide a 10-fold increase in resolution, allowing us to map the jet dynamics and evolving magnetic field structures closer to the central engine than ever before. Multi-frequency phase transfer and multi-frequency imaging will also enable higher image dynamic range, revealing detailed internal structure and dynamics within the jets \citep[e.g.,][]{Lobanov_2001,Gomez_2016,Gomez_2022,Okino_2022,Chael_2023_multifreq,2023NatAs...7.1359F}, including regions 10 times closer to the SMBH than is possible at cm wavelengths because of synchrotron self-absorption. These studies will also have multi-wavelength and multi-messenger components, with the potential to study the connection to high energy emission, including neutrinos.


\appendix

\begin{table}[t]
    \centering
    {
    \setlength{\fboxsep}{0pt}
    \fbox{\includegraphics[width=\textwidth]{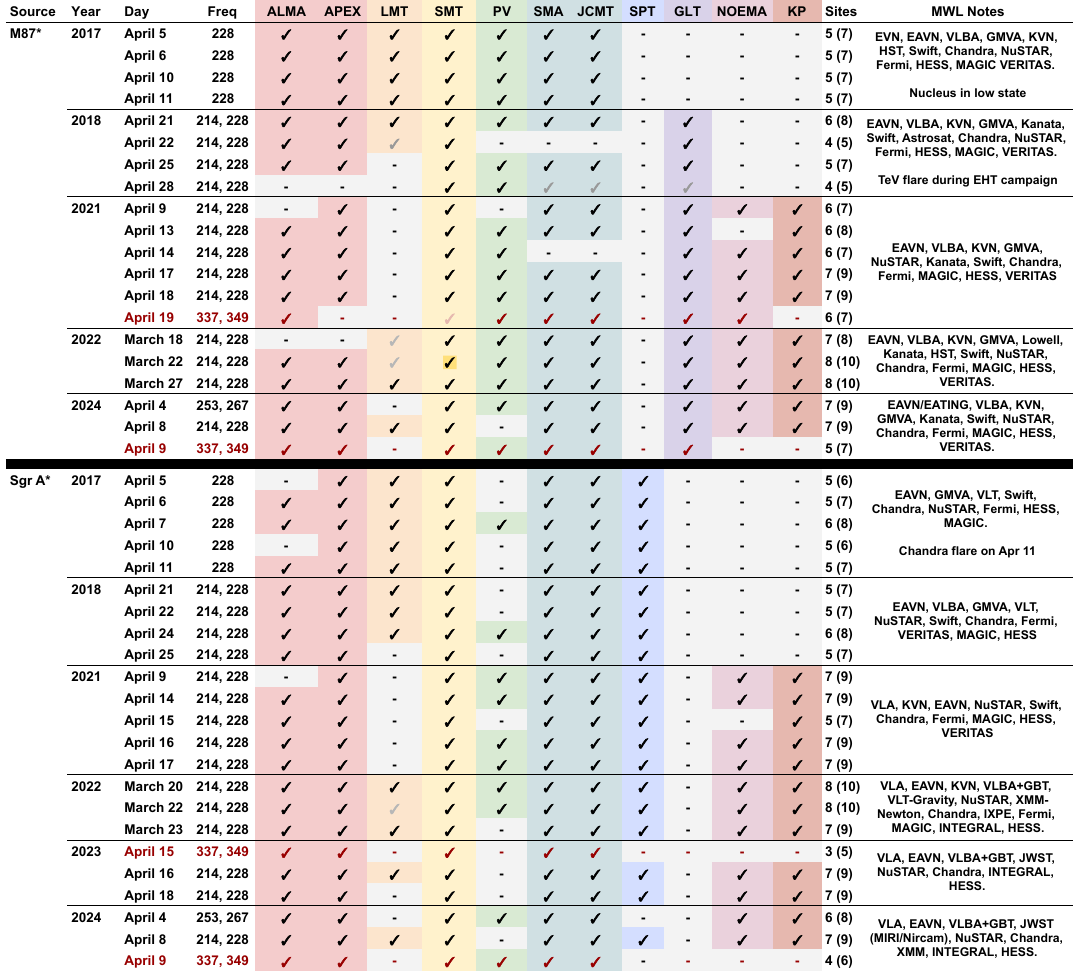}}\vspace{-0.4cm}
    }
    \caption{Summary of existing EHT observations of \m87 (top) and \sgra (bottom) from $2017-2024$, with participating observatories noted with a {\checkmark} ({\color{lightgray} \checkmark} indicates that participation was severely limited by weather or technical difficulties). Observations of \m87 and \sgra have primarily been conducted in the 230\,GHz band with one observation for each source at 345\,GHz. Sites joining these 345\,GHz observations are marked with a { \color{darkred} \checkmark}; the SMT for 2021 \m87 observations is denoted using a faint marker because it only joined for one partial and one full scan because of bad weather.  
    The ``Sites'' column lists the total number of geographic locations, with the total number of participating telescopes in parenthesis. 
    The EHT operations and analysis process are discussed in Section~\ref{sec:recommendations}.  KVN Yonsei observations with the EHT at 230 GHz were performed on a best-efforts basis for the first time in 2024.
    }
    \label{tab:eht_observations}
\end{table}

\section{EHT Observing Campaign Data Summary}

In this appendix, we summarize the existing EHT observations (Table~\ref{tab:eht_observations}) and describe the data handling process and associated timelines.  The table details all observations of \m87 and \sgra, including the antennas involved, the MWL resources used as part of the campaign, and the release date of basic data products.  

The EHT operations and science utilization typically span a multi-year cycle, from the identification of the year-specific science goals to the completion of science utilization and the announcement of results.
 Observations are proposed a year before the planned observing epoch. Assembling the full set of recorded data at the correlator sites takes nearly a year due to the limits on shipping from the South Pole station.  
Fringe-finding, production correlation, metadata collection, calibration, and data validation typically span an additional year. The duration of scientific analysis and paper writing is highly variable dependent on the complexity of the project, collaboration priorities, and available resources. Data issues can appear at this stage, requiring re-calibration and sometimes re-correlation. Data issues most commonly occur as the result of the introduction of new stations or new instrumental capabilities into the array.

Raw visibility data and associated metadata are posted to the ALMA data archive 1 year after they have passed internal validation. Calibrated visibility data and accompanying analysis routines are posted to the Cyverse archive at the time of publication. All archival data are available on the EHT public website:  \url{https://eventhorizontelescope.org/for-astronomers/data}.

\bibliography{references,osg}

\begin{CJK*}{UTF8}{gbsn}
\allauthors

\section*{Acknowledgements}
The Event Horizon Telescope Collaboration thanks the following
organizations and programs: the Academia Sinica; the Academy
of Finland (projects 274477, 284495, 312496, 315721); the Agencia Nacional de Investigaci\'{o}n 
y Desarrollo (ANID), Chile via NCN$19\_058$ (TITANs), Fondecyt 1221421 and BASAL FB210003; the Alexander
von Humboldt Stiftung; an Alfred P. Sloan Research Fellowship;
Allegro, the European ALMA Regional Centre node in the Netherlands, the NL astronomy
research network NOVA and the astronomy institutes of the University of Amsterdam, Leiden University, and Radboud University;
the ALMA North America Development Fund; the Astrophysics and High Energy Physics programme by MCIN (with funding from European Union NextGenerationEU, PRTR-C17I1); the Black Hole Initiative, which is funded by grants from the John Templeton Foundation (60477, 61497, 62286) and the Gordon and Betty Moore Foundation (Grant GBMF-8273) - although the opinions expressed in this work are those of the author and do not necessarily reflect the views of these Foundations; 
the Brinson Foundation; ``la Caixa'' Foundation (ID 100010434) through fellowship codes LCF/BQ/DI22/11940027 and LCF/BQ/DI22/11940030; 
Chandra DD7-18089X and TM6-17006X; the China Scholarship
Council; the China Postdoctoral Science Foundation fellowships (2020M671266, 2022M712084); Conicyt through Fondecyt Postdoctorado (project 3220195); Consejo Nacional de Humanidades, Ciencia y Tecnología (CONAHCYT, Mexico, projects U0004-246083, U0004-259839, F0003-272050, M0037-279006, F0003-281692, 104497, 275201, 263356, CBF2023-2024-1102, 257435); the Colfuturo Scholarship; 
the Consejer\'{i}a de Econom\'{i}a, Conocimiento, 
Empresas y Universidad 
of the Junta de Andaluc\'{i}a (grant P18-FR-1769), the Consejo Superior de Investigaciones 
Cient\'{i}ficas (grant 2019AEP112);
the Delaney Family via the Delaney Family John A.
Wheeler Chair at Perimeter Institute; Dirección General de Asuntos del Personal Académico-Universidad Nacional Autónoma de México (DGAPA-UNAM, projects IN112820 and IN108324); the Dutch Research Council (NWO) for the VICI award (grant 639.043.513), the grant OCENW.KLEIN.113, and the Dutch Black Hole Consortium (with project No. NWA 1292.19.202) of the research programme the National Science Agenda; the Dutch National Supercomputers, Cartesius and Snellius  
(NWO grant 2021.013); 
the EACOA Fellowship awarded by the East Asia Core
Observatories Association, which consists of the Academia Sinica Institute of Astronomy and
Astrophysics, the National Astronomical Observatory of Japan, Center for Astronomical Mega-Science,
Chinese Academy of Sciences, and the Korea Astronomy and Space Science Institute; 
the European Research Council (ERC) Synergy Grant ``BlackHoleCam: Imaging the Event Horizon of Black Holes'' (grant 610058) and Synergy Grant ``BlackHolistic:  Colour Movies of Black Holes:
Understanding Black Hole Astrophysics from the Event Horizon to Galactic Scales'' (grant 10107164); 
the European Union Horizon 2020
research and innovation programme under grant agreements
RadioNet (No. 730562), 
M2FINDERS (No. 101018682) and FunFiCO (No. 777740); the European Research Council for advanced grant `JETSET: Launching, propagation and 
emission of relativistic jets from binary mergers and across mass scales' (grant No. 884631); the European Horizon Europe staff exchange (SE) programme HORIZON-MSCA-2021-SE-01 grant NewFunFiCO (No. 10108625); the Horizon ERC Grants 2021 programme under grant agreement No. 101040021; the FAPESP (Funda\c{c}\~ao de Amparo \'a Pesquisa do Estado de S\~ao Paulo) under grant 2021/01183-8; the Fondo CAS-ANID folio CAS220010; 
the Generalitat
Valenciana (grants APOSTD/2018/177 and  ASFAE/2022/018) and
GenT Program (project CIDEGENT/2018/021); the Gordon and Betty Moore Foundation (GBMF-3561, GBMF-5278, GBMF-10423);   
the Institute for Advanced Study; the ICSC – Centro Nazionale di Ricerca in High Performance Computing, Big Data and Quantum Computing, funded by European Union – NextGenerationEU; the Istituto Nazionale di Fisica
Nucleare (INFN) sezione di Napoli, iniziative specifiche
TEONGRAV; 
the International Max Planck Research
School for Astronomy and Astrophysics at the
Universities of Bonn and Cologne; the Italian Ministry of University and Research (MUR)– Project CUP F53D23001260001, funded by the European Union – NextGenerationEU; 
DFG research grant ``Jet physics on horizon scales and beyond'' (grant No. 443220636);
Joint Columbia/Flatiron Postdoctoral Fellowship (research at the Flatiron Institute is supported by the Simons Foundation); 
the Japan Ministry of Education, Culture, Sports, Science and Technology (MEXT; grant JPMXP1020200109); 
the Japan Society for the Promotion of Science (JSPS) Grant-in-Aid for JSPS
Research Fellowship (JP17J08829); the Joint Institute for Computational Fundamental Science, Japan; the Key Research
Program of Frontier Sciences, Chinese Academy of
Sciences (CAS, grants QYZDJ-SSW-SLH057, QYZDJSSW-SYS008, ZDBS-LY-SLH011); 
the Leverhulme Trust Early Career Research
Fellowship; the Max-Planck-Gesellschaft (MPG);
the Max Planck Partner Group of the MPG and the
CAS; the MEXT/JSPS KAKENHI (grants 18KK0090, JP21H01137,
JP18H03721, JP18K13594, 18K03709, JP19K14761, 18H01245, 25120007, 19H01943, 21H01137, 21H04488, 22H00157, 23K03453); the MICINN Research Projects PID2019-108995GB-C22, PID2022-140888NB-C22; the MIT International Science
and Technology Initiatives (MISTI) Funds; 
the Ministry of Science and Technology (MOST) of Taiwan (103-2119-M-001-010-MY2, 105-2112-M-001-025-MY3, 105-2119-M-001-042, 106-2112-M-001-011, 106-2119-M-001-013, 106-2119-M-001-027, 106-2923-M-001-005, 107-2119-M-001-017, 107-2119-M-001-020, 107-2119-M-001-041, 107-2119-M-110-005, 107-2923-M-001-009, 108-2112-M-001-048, 108-2112-M-001-051, 108-2923-M-001-002, 109-2112-M-001-025, 109-2124-M-001-005, 109-2923-M-001-001, 
110-2112-M-001-033, 110-2124-M-001-007 and 110-2923-M-001-001); the National Science and Technology Council (NSTC) of Taiwan
(111-2124-M-001-005, 112-2124-M-001-014 and  112-2112-M-003-010-MY3);
the Ministry of Education (MoE) of Taiwan Yushan Young Scholar Program;
the Physics Division, National Center for Theoretical Sciences of Taiwan;
the National Aeronautics and
Space Administration (NASA, Fermi Guest Investigator
grant 
80NSSC23K1508, NASA Astrophysics Theory Program grant 80NSSC20K0527, NASA NuSTAR award 
80NSSC20K0645); NASA Hubble Fellowship Program Einstein Fellowship;
NASA Hubble Fellowship 
grants HST-HF2-51431.001-A, HST-HF2-51482.001-A, HST-HF2-51539.001-A awarded 
by the Space Telescope Science Institute, which is operated by the Association of Universities for 
Research in Astronomy, Inc., for NASA, under contract NAS5-26555; 
the National Institute of Natural Sciences (NINS) of Japan; the National
Key Research and Development Program of China
(grant 2016YFA0400704, 2017YFA0402703, 2016YFA0400702); the National Science and Technology Council (NSTC, grants NSTC 111-2112-M-001 -041, NSTC 111-2124-M-001-005, NSTC 112-2124-M-001-014); the US National
Science Foundation (NSF, grants AST-0096454,
AST-0352953, AST-0521233, AST-0705062, AST-0905844, AST-0922984, AST-1126433, OIA-1126433, AST-1140030,
DGE-1144085, AST-1207704, AST-1207730, AST-1207752, MRI-1228509, OPP-1248097, AST-1310896, AST-1440254, 
AST-1555365, AST-1614868, AST-1615796, AST-1715061, AST-1716327,  AST-1726637, 
OISE-1743747, AST-1743747, AST-1816420, AST-1935980, AST-1952099, AST-2034306,  AST-2205908, AST-2307887); 
NSF Astronomy and Astrophysics Postdoctoral Fellowship (AST-1903847); 
the Natural Science Foundation of China (grants 11650110427, 10625314, 11721303, 11725312, 11873028, 11933007, 11991052, 11991053, 12192220, 12192223, 12273022, 12325302, 12303021); 
the Natural Sciences and Engineering Research Council of
Canada (NSERC, including a Discovery Grant and
the NSERC Alexander Graham Bell Canada Graduate
Scholarships-Doctoral Program); 
the National Research Foundation of Korea (the Global PhD Fellowship Grant: grants NRF-2015H1A2A1033752; the Korea Research Fellowship Program: NRF-2015H1D3A1066561; Brain Pool Program: RS-2024-00407499;  Basic Research Support Grant 2019R1F1A1059721, 2021R1A6A3A01086420, 2022R1C1C1005255, 2022R1F1A1075115); 
Netherlands Research School for Astronomy (NOVA) Virtual Institute of Accretion (VIA) postdoctoral fellowships; NOIRLab, which is managed by the Association of Universities for Research in Astronomy (AURA) under a cooperative agreement with the National Science Foundation; 
Onsala Space Observatory (OSO) national infrastructure, for the provisioning
of its facilities/observational support (OSO receives
funding through the Swedish Research Council under
grant 2017-00648);  the Perimeter Institute for Theoretical
Physics (research at Perimeter Institute is supported
by the Government of Canada through the Department
of Innovation, Science and Economic Development
and by the Province of Ontario through the
Ministry of Research, Innovation and Science); the Portuguese Foundation for Science and Technology (FCT) grants (Individual CEEC program - 5th edition, \url{https://doi.org/10.54499/UIDB/04106/2020}, \url{https://doi.org/10.54499/UIDP/04106/2020}, PTDC/FIS-AST/3041/2020, CERN/FIS-PAR/0024/2021, 2022.04560.PTDC); the Princeton Gravity Initiative; the Spanish Ministerio de Ciencia e Innovaci\'{o}n (grants PGC2018-098915-B-C21, AYA2016-80889-P,
PID2019-108995GB-C21, PID2020-117404GB-C21); 
the University of Pretoria for financial aid in the provision of the new 
Cluster Server nodes and SuperMicro (USA) for a SEEDING GRANT approved toward these 
nodes in 2020; the Shanghai Municipality orientation program of basic research for international scientists (grant no. 22JC1410600); 
the Shanghai Pilot Program for Basic Research, Chinese Academy of Science, 
Shanghai Branch (JCYJ-SHFY-2021-013); the Simons Foundation (grant 00001470);
the State Agency for Research of the Spanish MCIU through
the ``Center of Excellence Severo Ochoa'' award for
the Instituto de Astrof\'{i}sica de Andaluc\'{i}a (SEV-2017-
0709); the Spanish Ministry for Science and Innovation grant CEX2021-001131-S funded by MCIN/AEI/10.13039/501100011033; the Spinoza Prize SPI 78-409; the South African Research Chairs Initiative, through the 
South African Radio Astronomy Observatory (SARAO, grant ID 77948),  which is a facility of the National 
Research Foundation (NRF), an agency of the Department of Science and Innovation (DSI) of South Africa; the Swedish Research Council (VR); the Taplin Fellowship; the Toray Science Foundation; the UK Science and Technology Facilities Council (grant no. ST/X508329/1); the US Department of Energy (USDOE) through the Los Alamos National
Laboratory (operated by Triad National Security,
LLC, for the National Nuclear Security Administration
of the USDOE, contract 89233218CNA000001); and the YCAA Prize Postdoctoral Fellowship. This work was supported by the National Research Foundation of Korea(NRF) grant funded by the Korea government(MSIT) (RS-2024-00449206).

We thank
the staff at the participating observatories, correlation
centers, and institutions for their enthusiastic support.
This paper makes use of the following ALMA data:
ADS/JAO.ALMA\#2016.1.01154.V,
 ADS/JAO.ALMA\#2016.1.01404.V,
 ADS/JAO.ALMA\#2017.1.00797.V,
 ADS/JAO.ALMA\#2017.1.00841.V,
 ADS/JAO.ALMA\#2018.1.01159.V,
 ADS/JAO.ALMA\#2019.1.01797.V,
 ADS/JAO.ALMA\#2019.1.01812.V,
 ADS/JAO.ALMA\#2021.1.00906.V, 
 ADS/JAO.ALMA\#2021.1.00910.V,
 ADS/JAO.ALMA\#2022.1.01268.V,
 ADS/JAO.ALMA\#2023.1.01243.V,
 and
 ADS/JAO.ALMA\#2023.1.01244.V.
ALMA is a partnership
of the European Southern Observatory (ESO;
Europe, representing its member states), NSF, and
National Institutes of Natural Sciences of Japan, together
with National Research Council (Canada), Ministry
of Science and Technology (MOST; Taiwan),
Academia Sinica Institute of Astronomy and Astrophysics
(ASIAA; Taiwan), and Korea Astronomy and
Space Science Institute (KASI; Republic of Korea), in
cooperation with the Republic of Chile. The Joint
ALMA Observatory is operated by ESO, Associated
Universities, Inc. (AUI)/NRAO, and the National Astronomical
Observatory of Japan (NAOJ). The NRAO
is a facility of the NSF operated under cooperative agreement
by AUI.
This research used resources of the Oak Ridge Leadership Computing Facility at the Oak Ridge National
Laboratory, which is supported by the Office of Science of the U.S. Department of Energy under contract
No. DE-AC05-00OR22725; the ASTROVIVES FEDER infrastructure, with project code IDIFEDER-2021-086; the computing cluster of Shanghai VLBI correlator supported by the Special Fund 
for Astronomy from the Ministry of Finance in China;  
We also thank the Center for Computational Astrophysics, National Astronomical Observatory of Japan. This work was supported by FAPESP (Fundacao de Amparo a Pesquisa do Estado de Sao Paulo) under grant 2021/01183-8.

APEX is a collaboration between the
Max-Planck-Institut f{\"u}r Radioastronomie (Germany),
ESO, and the Onsala Space Observatory (Sweden). The
SMA is a joint project between the SAO and ASIAA
and is funded by the Smithsonian Institution and the
Academia Sinica. The JCMT is operated by the East
Asian Observatory on behalf of the NAOJ, ASIAA, and
KASI, as well as the Ministry of Finance of China, Chinese
Academy of Sciences, and the National Key Research and Development
Program (No. 2017YFA0402700) of China
and Natural Science Foundation of China grant 11873028.
Additional funding support for the JCMT is provided by the Science
and Technologies Facility Council (UK) and participating
universities in the UK and Canada. 
The LMT is a project operated by the Instituto Nacional
de Astr\'{o}fisica, \'{O}ptica, y Electr\'{o}nica (Mexico) and the
University of Massachusetts at Amherst (USA). The
IRAM 30-m telescope on Pico Veleta, Spain is operated
by IRAM and supported by CNRS (Centre National de
la Recherche Scientifique, France), MPG (Max-Planck-Gesellschaft, Germany), 
and IGN (Instituto Geogr\'{a}fico
Nacional, Spain). The SMT is operated by the Arizona
Radio Observatory, a part of the Steward Observatory
of the University of Arizona, with financial support of
operations from the State of Arizona and financial support
for instrumentation development from the NSF.
Support for SPT participation in the EHT is provided by the National Science Foundation through award OPP-1852617 
to the University of Chicago. Partial support is also 
provided by the Kavli Institute of Cosmological Physics at the University of Chicago. The SPT hydrogen maser was 
provided on loan from the GLT, courtesy of ASIAA.

This work used the
Extreme Science and Engineering Discovery Environment
(XSEDE), supported by NSF grant ACI-1548562,
and CyVerse, supported by NSF grants DBI-0735191,
DBI-1265383, and DBI-1743442. XSEDE Stampede2 resource
at TACC was allocated through TG-AST170024
and TG-AST080026N. XSEDE JetStream resource at
PTI and TACC was allocated through AST170028.
This research is part of the Frontera computing project at the Texas Advanced 
Computing Center through the Frontera Large-Scale Community Partnerships allocation
AST20023. Frontera is made possible by National Science Foundation award OAC-1818253.
This research was done using services provided by the OSG Consortium~\citep{osg07,osg09}, which is supported by the National Science Foundation award Nos. 2030508 and 1836650.
Additional work used ABACUS2.0, which is part of the eScience center at Southern Denmark University, and the Kultrun Astronomy Hybrid Cluster (projects Conicyt Programa de Astronomia Fondo Quimal QUIMAL170001, Conicyt PIA ACT172033, Fondecyt Iniciacion 11170268, Quimal 220002). 
Simulations were also performed on the SuperMUC cluster at the LRZ in Garching, 
on the LOEWE cluster in CSC in Frankfurt, on the HazelHen cluster at the HLRS in Stuttgart, 
and on the Pi2.0 and Siyuan Mark-I at Shanghai Jiao Tong University.
The computer resources of the Finnish IT Center for Science (CSC) and the Finnish Computing 
Competence Infrastructure (FCCI) project are acknowledged. This
research was enabled in part by support provided
by Compute Ontario (http://computeontario.ca), Calcul
Quebec (http://www.calculquebec.ca), and Compute
Canada (http://www.computecanada.ca). 

The EHTC has
received generous donations of FPGA chips from Xilinx
Inc., under the Xilinx University Program. The EHTC
has benefited from technology shared under open-source
license by the Collaboration for Astronomy Signal Processing
and Electronics Research (CASPER). The EHT
project is grateful to T4Science and Microsemi for their
assistance with hydrogen masers. This research has
made use of NASA's Astrophysics Data System. We
gratefully acknowledge the support provided by the extended
staff of the ALMA, from the inception of
the ALMA Phasing Project through the observational
campaigns of 2017 and 2018. We would like to thank
A. Deller and W. Brisken for EHT-specific support with
the use of DiFX. We thank Martin Shepherd for the addition of extra features in the Difmap software 
that were used for the CLEAN imaging results presented in this paper.
We acknowledge the significance that
Maunakea, where the SMA and JCMT EHT stations
are located, has for the indigenous Hawaiian people.

\end{CJK*}

\end{document}